\documentclass[a4paper]{article}

\usepackage{INTERSPEECH2020}
\usepackage{multirow}
\usepackage[noend]{algpseudocode}
\usepackage{algorithmicx,algorithm}
\usepackage{color,xcolor}
\usepackage{caption}
\usepackage{graphicx}
\usepackage{makecell}
\usepackage{bbding}
\usepackage{pifont}
\usepackage{wasysym}
\usepackage{amssymb}
\usepackage[colorlinks,linkcolor=blue,]{hyperref}
\usepackage{subfigure}
\title{Channel-wise Subband Input for Better Voice and Accompaniment Separation on High Resolution Music}
\name{Haohe Liu, Lei Xie$^*$, Jian Wu, Geng Yang}
\address{
	Audio, Speech and Language Processing Group (ASLP@NPU),
	School of Computer Science, Northwestern Polytechnical University, Xi’an, China
}
\email{haoheliu@gmail.com, \{lxie,jianwu,gengyang\}@nwpu-aslp.org}

\newcommand{\tabincell}[2]{\begin{tabular}{@{}#1@{}}#2\end{tabular}}  

\begin{document}
	
	\maketitle
	\begin{abstract}
		This paper presents a new input format, channel-wise subband input (CWS), for convolutional neural networks (CNN) based 
		music source separation (MSS) models in the frequency domain. 
		We aim to address the major issues in CNN-based high-resolution MSS model: high computational cost and weight sharing between distinctly different bands. 
		Specifically, in this paper, we decompose the input mixture spectra into several bands and concatenate them channel-wise as the model input. 
		The proposed approach enables effective weight sharing in each subband and introduces more flexibility between channels.  
		For comparison purposes, we perform voice and accompaniment separation (VAS) on models with different scales, 
		architectures, and CWS settings. Experiments show that the CWS input is beneficial in many aspects. We evaluate our method on \textit{musdb18hq} test set, focusing on SDR, SIR and SAR metrics. Among all our experiments, 
		CWS enables models to obtain 6.9\% performance gain on the average metrics. With even a smaller number of parameters, 
		less training data, and shorter training time, our MDenseNet with 8-bands CWS input still surpasses the original MMDenseNet with a large margin. 
		Moreover, CWS also reduces computational cost and training time to a large extent.
		%    Most frequency based music source separation models take the full band spectrogram as the input feature.
		
		%    This paper presents a new input format, channel-wise subband input (CW1S), for CNN based high-resolution music source separation models in the frequency domain. Current deep-learning based models usually suffer from high computational cost and long training time. We summarize the weakness of full-band input in CNN architecture. We aim to reduce computational cost and boosting performance by utilizing features in different bands. For comparison purpose, we perform experiments on models with different scales, architectures as well as CWS settings. The result shows 4-channels CWS input enables models to obtain a 5\% performance gain on average, less calculation, and much shorter training time.  
	\end{abstract}
	
	\let\thefootnote\relax\footnotetext{*Corresponding Author}
	
	\section{Introduction}
	
	Music Source Separation (MSS) has raised much interest in recent years. 
	The goal of the task is blindly separate sources from a mixed track, 
	for example vocal, drums, bass and accompaniment. In this paper, 
	we particularly focus on the voice and accompaniment separation (VAS) from a mixture. 
	As a practical tool, separating these two components allows us to remix, 
	suppress or up-mix the sources~\cite{cano2018musical}. VAS can also facilitate automatic 
	transcription, karaoke track generating as well as music information retrieval~\cite{perez2019improving}.
	
	High-resolution music usually sounds better but suffers from high computational 
	cost in the VAS task. For example, 44.1kHz is a commonly used sample rate for music, while many high-quality formats may be up to 48kHz or even higher. However, due to the high-computational cost, many of the current VAS studies perform downsampling in advance. For instance, the approach using M-U-Net~\cite{kadandale2020multi} downsamples the audio to 10.88kHz before processing and Dense-Unet only works on 16kHz music in ~\cite{liu2020voice}. The downsampling process seriously affects the auditory quality to the separated vocal and accompaniment in practical applications.
	
%	This kind of music has better resolution thus sound finer 
%	and smoother. But the sample rate is not the higher the better. 
%	According to the perceptible frequency range of human ears and Nyquist theory, 
%	the upper bound is approximately 40kHz. Thus we use 44.1 kHz in our work, 
%	which is also the standard sample rate for music. However, 44.1 kHz is not a widely 
%	chosen sample rate due to the high cost of computation. To expedite the training 
%	process, some works do the downsampling in advance, e.g.,  But the auditory quality they produce is not as good as 
%	model trained with 44.1 kHz data.
	
	Convolutional Neural Networks (CNN) has shown tremendous success in multiple fields, 
	especially image-related tasks. The input data for these tasks, such as image classification, usually 
	have problem that the position of a certain object is not fixed. Mechanisms like local receptive fields and 
	shared weights~\cite{10.5555/303568.303704} enable CNN to 
	become position invariant, which means once a feature has been detected, 
	its exact location becomes less important~\cite{10.5555/303568.303704}. In audio processing, most of the state-of-the-art (SOTA) 
	MSS models are also based on convolutional networks, like Deep-Unet~\cite{jansson2017singing}, which 
	have shown considerable improvements over the traditional methods.
	
\begin{figure}[tbh]
		\centering
		\includegraphics[width=0.7\linewidth]{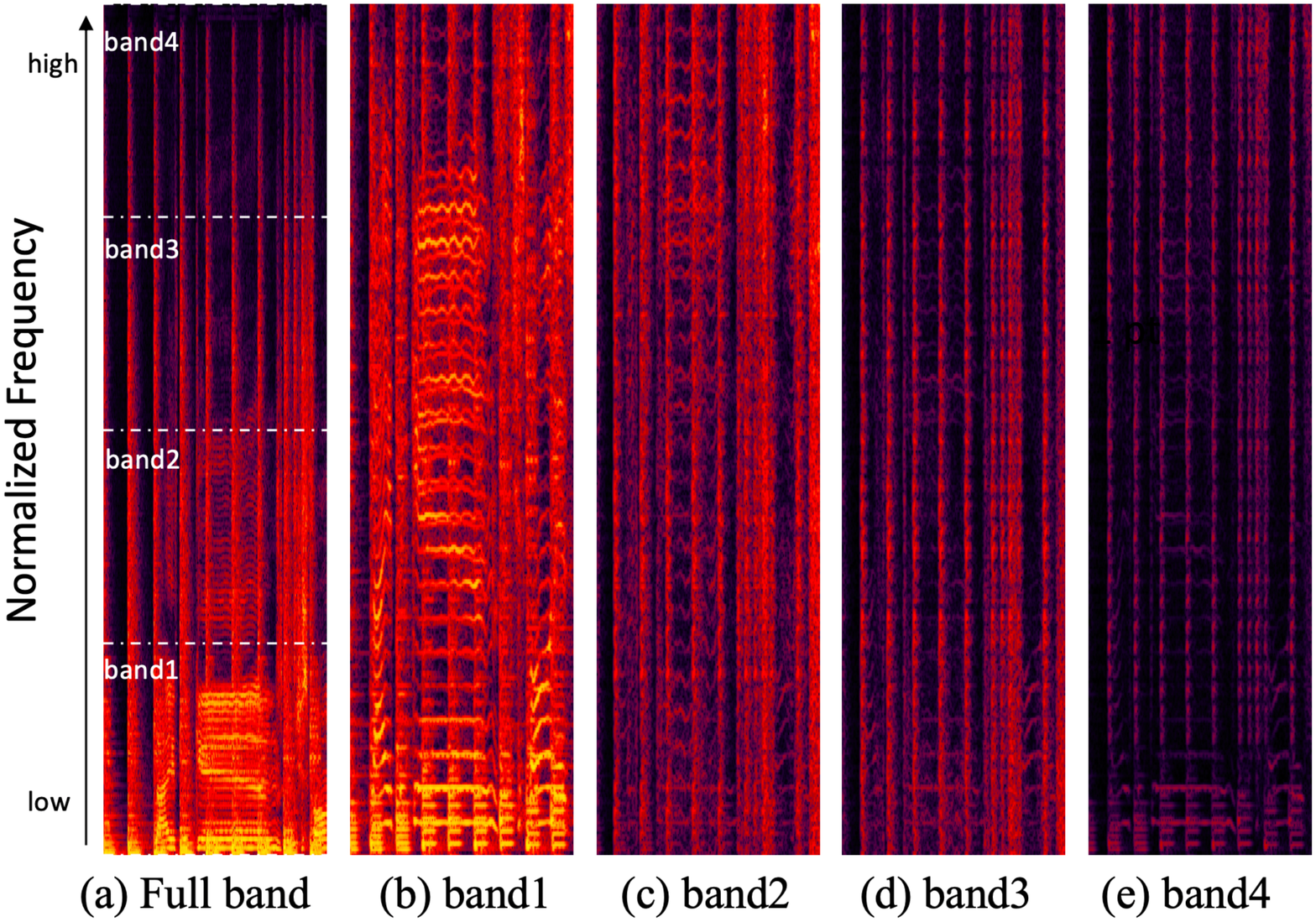}
		\caption{Comparison between different bands. Lower frequency band contains more energy, 
			long sustained sound, fundamental frequency and harmonic series, 
			while higher band, mostly percussive signals and low-energy resonance, 
			contains less energy and less complex information.}
		\label{fig:subbands}
	\end{figure}

	Although CNN-based architecture has demonstrated effectiveness on MSS tasks 
	in frequency domain, it still has apparent limitations. Frequency spectrogram based 
	SOTA models trained on high-resolution audio, e.g. 
	TFC-TIF~\cite{choi2019investigating}, 
	% V1.2 
	%and MMDenseNet~\cite{takahashi2017multi}
	usually take the whole 
    spectrogram as the input feature. In this case, they assume each frequency 
	band has filter parameters to share and are equally important. However, 
	local patterns are usually different between bands~\cite{takahashi2017multi}, as can be seen in Fig~\ref{fig:subbands}.
	This means different bands do not necessarily need the same set of filters (in CNN) 
	for parameter sharing. 
%	Besides, the lower band contains more energy than the higher band. 
	Hence, treating different frequency bands differently might better
	facilitate the separation process.
	
	% the lower frequency band is more likely to contain high energies, tonalities, and long sustained sounds, whereas the higher frequency band tends to have low energies, noise, and rapidly decaying sounds 
	%after statistic of the training set, we found distinctly different patterns between frequency bands. First, the magnitude feature itself is different. Lower frequency band usually consists of the fundamental frequency, multiple resonance humps, and other complex patterns. While higher band, mostly percus1sive signals and low-energy resonance peaks, contains less complex information.
	
	Some prior efforts have already emphasized the difference between bands. In ~\cite{taghia2009subband}, 
	Taghia \textit{et al.} first took the subband decomposition, and then they used a hybrid system 
	of empirical mode decomposition~\cite{huang1998the} and principle component analysis to 
	construct artificial observations from the single mixture. Finally a synthesis process 
	was used to reconstruct full band signal. Takahashi \textit{et al.} ~\cite{takahashi2018mmdenselstm} also noticed the 
	problem of global kernel sharing. They pointed out that the global weight sharing 
	works well on natural photos, in which local pattern appears in any position of 
	the input~\cite{takahashi2017multi}. But this is not the case for audio. To handle this problem, 
	they designed the dedicated MDenseNets for each frequency bands and a full band 
	MDenseNet for full band rough structure~\cite{takahashi2017multi}. This model 
	achieved the state-of-the-art performance 
	on SiSEC 2016 competition~\cite{liutkus2017the}. 
	%In SiSEC 2018~\cite{Stoter2018}, MMDenseLSTM adapted that same scheme, outperform MMDenseNet and achieved b results.
	 
	In this paper, we propose a new input format for the MSS model in the frequency domain, namely channel-wise subband (CWS) input. Different from the band-dedicated approach in ~\cite{takahashi2017multi,takahashi2018mmdenselstm}, our method can handle both sub-bands and full-band in a single model, which makes CWS-based model highly efficient, less complex, and easier to use.  We extensively evaluate our method on MDenseNet, UNet~\cite{jansson2017singing} 
	with different scales, and three kinds of subband settings on \textit{musdb18hq}~\cite{musdb18-hq}. 
	We also test our approach on a larger internal dataset. Results 
	show that the model with CWS input not only outperforms 
	the model without CWS by a large margin, but also boosts the speed of model training 
	as well as separation. 
	
	\section{Methodology}
	\begin{figure}[b]
		\centering
		\includegraphics[width=1.0\linewidth]{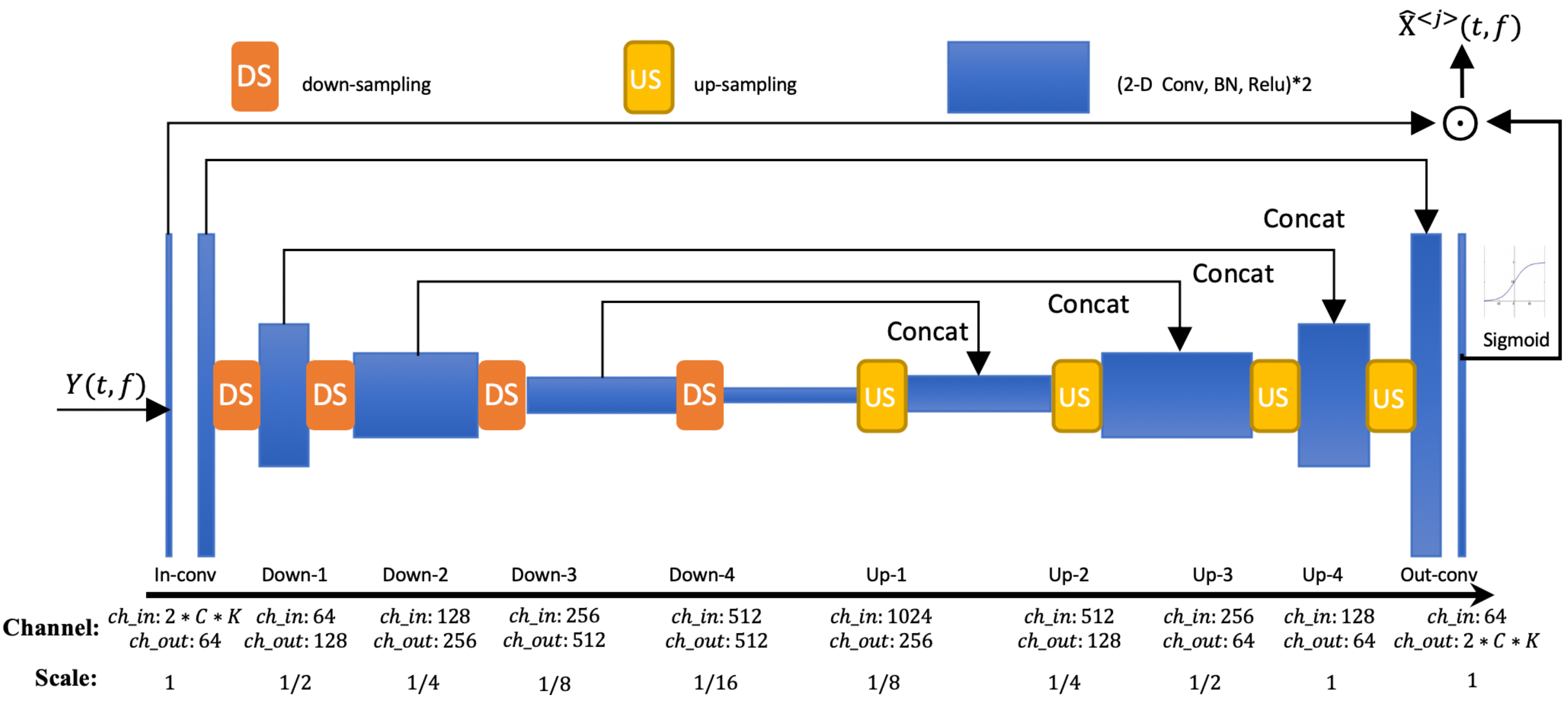}
		\caption{Unet Structure used in this paper}
		\label{fig:screenshot001}
	\end{figure}
	
	Given the raw music signal $y(n)$, our goal is to separate a set of source 
	signals $x(n)=\{x_j(n)\}, j \in \{1, \cdots, J\}$. In this paper, we focus on 
	VAS task. Thus $J = 2$ and $x_1(n), x_2(n)$ are vocal and accompaniment track, respectively. 
	In the time domain, the observed mixture signal is modeled as:
	\begin{equation}
	y(n) = x_1(n) + x_2(n),
	\end{equation}
	and in the frequency domain, it equals to
	\begin{equation}
	\label{energy-conserv}
	Y(t, f) = X_1(t, f) + X_2(t, f),
	\end{equation}
	where $t, f$ index time and frequency axis. $Y, X_1, X_2$ are short-time fourier 
	transform (STFT) of the mixture signal $y(n)$ and the source signal $x_1(n), x_2(n)$.
	Also we should clarify that $Y$ 
	is not a magnitude spectrogram as was used in conventional frequency domain models~\cite{chandna2020content,jansson2017singing,liu2020voice,uhlich2017improving}, but the complex-valued STFT matrices. This method is explored by ~\cite{choi2019investigating} 
	with considerable improvements. 
	% hathon
	In this section, we will briefly introduce the UNet, MMDenseNet, the proposed analysis-synthesis scheme and CWS feature format. 
	
	%After that we describe the details of our model training process
	
	\subsection{UNet}
	Fig~\ref{fig:screenshot001} depicts the structure of UNet~\cite{ronneberger2015u-net:} we use in this paper. The number of different scales $s=5$. It takes the mixture spectra $Y(t,f)$ as input and outputs 
	the Time-Frequency Mask~\cite{huang2014deep} $M_j(t,f)$ for source $j$, which has identical size 
	with the input. The \textit{in-conv} block firstly expands the input channel to 64. 
	After that, we go through a series of convolution blocks, down/upsampling layers with 
	skip connections. Each convolution block consists of two series of 2D convolution layer, 
	batch normalization and rectified linear units~\cite{glorot2011deep}. We use $3\times3$ kernel size in 
	convolution layers, with the padding value of 1 to make sure that the frequency and 
	time dimension will not be changed by the convolution operations. 
	We use max-pooling and linear interpolation to scale down and up the feature map 
	by a factor of 2. Skip connections are added between down and up path. 
	The input feature of each convolution layer in the up path is concatenated 
	with the same scale output in the down path. To constraint the mask value 
	between $[0, 1]$, the sigmoid function is added on the model's output. Finally 
	the source estimation is obtained by multiply the mask and the mixture 
	STFT: 
	\begin{equation}
	\label{model_def}
	\hat{X}_j(t,f) = M_j(t,f) \cdot Y_j(t,f).
	\end{equation}
	The final separated music signal $\hat{x}_j(n)$ is obtained through inverse short-time fourier 
	transform (iSTFT) using the source estimation in Eq.~(\ref{model_def}).
	% In our work we also test unet with $s=6$. The basic structure is similar between different scale unets. 
	
	\begin{table}[tbh]
		\centering
		\setlength{\belowcaptionskip}{10pt} 
		\setlength{\abovecaptionskip}{-10pt} 
		\includegraphics[width=0.7\linewidth]{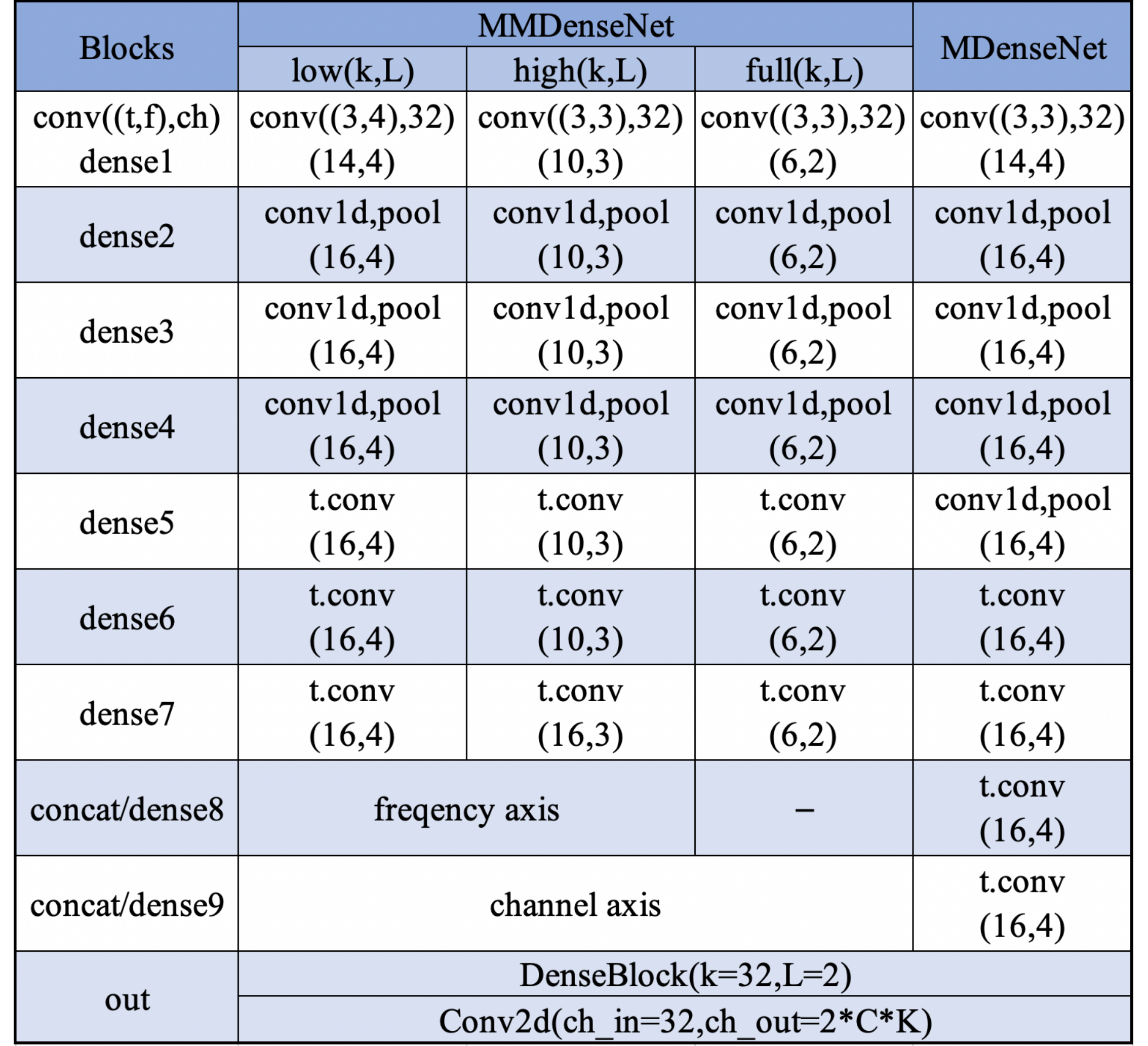}
		\caption{Details of MMDenseNet and MDenseNet.}
						\vspace{-5pt}
		\label{tab:MMDenseNet and MDenseNet}%
	\end{table}

	\subsection{MMDenseNet}
	With only about 0.3M parameters and SOTA performance, MMDenseNet~\cite{takahashi2017multi}
	is currently one of the most effective models for audio source separation. 
	It utilizes the characteristics in each frequency band to design different MDenseNets dedicated for each band. The frequency axis size of band-dedicated MDenseNet will be reduced to $f/K$ for it only processes data within that specific band.
	The outputs of these MDenseNets, $\hat{X}_{k}(t,f/K)$, are concatenated in 
	the frequency axis to recover full band prediction $\hat{X}_{\text{sub}}(t,f)$. 
	To capture the global rough structure, MMDenseNet also has a MDenseNet for the 
	full band. Then the output of the full band MDenseNet $\hat{X}_{\text{full}}(t,f)$, 
	is concatenated with $\hat{X}_{\text{sub}}(t,f)$ and pass through a final 
	denseblock to recover the final prediction $\hat{X}(t,f)$. Each MDenseNet 
	inside MMDenseNet can be designed independently according to its function 
	and data complexity.  MMDenseNet is highly parameter efficient due to the
	use of denseblocks~\cite{huang2017densely} and skip-connection between denseblocks. 
	Inside denseblock, the input of Denselayers is the concatenation of 
	previous layers' output or the skip connection of the former dense block. 
	This scheme enables model to reuse feature effectively, and thus it is highly 
	parameter efficient. 
	The MMDenseNet we use in this paper has a scale of 4, with three MDenseNets, as is shown in Table~\ref{tab:MMDenseNet and MDenseNet}. The detail of 
	MMDenseNet is described in ~\cite{takahashi2017multi}. 
	
	In this paper, we also conduct experiments on MDenseNet, both with CWS and 
	without CWS. The structure of MDenseNet we use is shown in 
	Table~\ref{tab:MMDenseNet and MDenseNet}. For a fair comparison with 
	MMDenseNet, we add two additional dense blocks to the lower part of 
	MMDenseNet to form the MDenseNet we use. In this way, the scale of our 
	MDenseNet is 5 and the total parameter number is 0.27 M.
	
	\subsection{Channel-wise Subband Input}
	
	% designed for audio downsampling and upsampling 
	We follow the method in ~\cite{yu2019durian:} for subband decomposing and signal reconstruction in the analysis and synthesis procedure. Both analysis and synthesis include a group of finite impulse response (FIR) uniform filter banks. We design three sets of analysis filter banks $H_{k}(e^{j\omega})$ and corresponding  
	synthesis filters $G_{k}(e^{j\omega})$, where $k\in{1,...,K}$ stands 
	for the number of subbands. The design of these filters follows the 
	procedure in ~\cite{Moazzen2014}. We use $y_{k}(n)$ to denote the output of 
	$H_{k}(e^{j\omega})$. After downsampling, the sample 
	rate of $y_{k}(n)$ is  $\frac{1}{K}$ of $y(n)$.
	
	%\begin{figure}[h]
	%    \centering
	%    \includegraphics[width=0.9\linewidth]{ana-syn}
	%    \caption{Channel-wise subband input}
	%    \label{fig:ana-syn}
	%\end{figure}
	
	\begin{figure}[tbh]
		\centering
				\setlength{\belowcaptionskip}{-6pt} 
		\setlength{\abovecaptionskip}{10pt} 
		\includegraphics[width=1\linewidth]{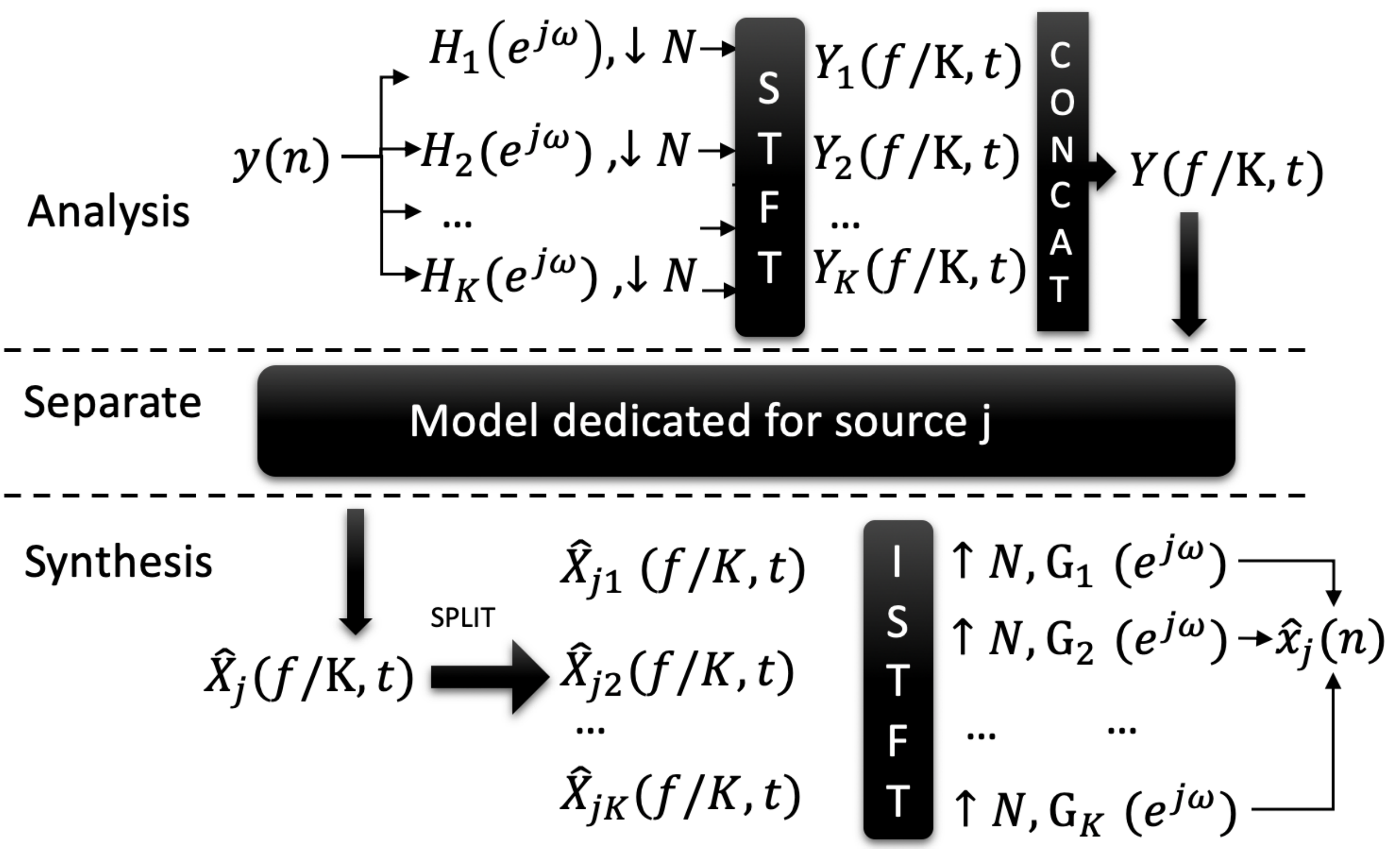}
		\caption{Channel-wise subband input}
		\label{fig:ana-syn}
	\end{figure}

	% The decrease of sample rate is significant for it greatly reduces the size of the input spectra. For example, for a three seconds 44.1KHz audio input, if we set the FFT size to 2048, window, and hop length to 32ms and 8ms, then the input feature map size will be $(channels,1024,376)$. But if we set $K=4$ in figure.~\ref{fig:ana-syn}, each analysis filter output will be subband signal sampled at 11.025kHz. In this case, the input feature map size will be (channels*K ,256,376), which will save about 75\% of memory usage and computational cost. 
	
	Though the total volume of input feature does not change, 
	the channel-wise concatenation of subbands is a better input format 
	for the frequency domain model. Here we give a simple explanation. 
	We use $\beta_{\mu}^{(l)}$ to denote the output feature map in $l$-th layer, $\mu$ 
	channel, and $S_{\lambda,\mu}^{(l-1)}$ to stand for the $\lambda$-th 
	convolution filters in layer $l-1$, which output is the $\mu$-th channel 
	in $l$-th layer. The 2D convolution layer can be 
	described as
	\begin{align}
		\label{cnn}
		o_{\mu}^{(l)}(i,j) & = \sum_{\lambda} \beta_{\lambda}^{(l-1)} * S_{\lambda,\mu}^{(l-1)} \\
		& = \sum_{\lambda} \sum_{m}\sum_{n}\beta_{\lambda}^{(l-1)}(i-m,j-n)S_{\lambda,\mu}^{(l-1)}(m,n) \notag \\
%	and \notag  \\
		\beta_{\mu}^{(l)}(i,j) & = \phi \left(o_{\mu}^{(l)}(i,j) \right).
	\end{align}
	
	From Eq.~(\ref{cnn}), we can observe that internal variable $o_{\mu}^{(l)}$ is 
	the linear product and sum of $\beta_{\lambda}^{(l-1)}$ and 
	$S_{\lambda,\mu}^{(l-1)}$. Thus we can not only view the convolutional kernel as 
	feature extractor, but also weight between different channels. For example, 
	if some filters in $S_{\mu}^{(l-1)}$ are set to all zero, then the corresponding 
	channels in $\beta_\mu^{(l-1)}$  will not be able to pass their value 
	to $\beta_{\mu}^{(l)}$. In this way, the $\mu$-th channel in feature 
	map $\beta^{(l)}$ can select the exact channels to be used in the previous 
	layer. The channel-wise concatenation enables model to assign different 
	capability on channel dimension, which is helpful to make the model highly efficient.
	
	After the analysis process, we perform STFT for each $y_{k}(n)$
	and the result is denoted as $Y_{k}(f/K,t)$.  
	% V1.2
	Here the sample rate of $y_{k}(n)$ reduce by a factor of $K$, $K \in \{2,4,8 \}$. So the 
	size of frequency axis will also be reduced $K$ fold. Then we concatenate $Y_{k}(f/K,t)$ along the subband dimension
	\begin{equation}
	Y=\left( \left[Y^{1}_{1},Y^{1}_{2},...,Y^{1}_{K},Y^{2}_{1},Y^{2}_{2},...,Y^{C}_{K}\right]\right)
	\label{Stack}
	\end{equation}
	to form the input feature of the network $Y(f/K,t)$. Since the data we use in this paper are all 
	stereo, $C$ equals to 2 here. The subscript $k$ indexes subband and we omit the 
	$(f/K,t)$ part in this equation for simplicity. 
	We treat different bands as different channels 
	so that model can both learn different channels independently and incorporate bands'
	features in a deeper layer. 
	
	% Consequently, $Y^{<j>}$ shape is $(2*C*K,f/K,t)$
	
	The synthesis procedure is a reverse version of the analysis. 
	We split the network output $\hat{X}_j(f/K,t)$ channel-wise as the prediction 
	of each subband. After iSTFT, we pass the result through a set of synthesis 
	filters to reconstruct source signal $\hat{x}_j(n)$.
	
	\subsection{Model Training}
	
	The synthesis procedure is not performed during training and the loss function is 
	defined as the sum of two components
	\begin{equation}
	\mathcal{L} = \mathcal{L}_1 + \mathcal{L}_c,
	\end{equation}
	where $\mathcal{L}_1$ is the $L_1$ norm and $\mathcal{L}_c$ denotes conservation 
	loss. Conservation loss could help when two dedicated models are trained jointly because it follows the basic model in Eq.~(\ref{energy-conserv}) and unites two independent dedicated-models. Each loss
	function measures the mean absolute error between network output $\hat{X}_j(f/K,t)$ 
	and the corresponding reference magnitude:
	\begin{equation}
	\begin{aligned}
	\mathcal{L}_1 & = \sum_{j=1}^{2} \sum_{t,f}  \left\vert\hat{X}_j(f/K,t)-X_j(f/K,t) \right\vert \\
	\mathcal{L}_c & = \sum_{j=1}^{2} \sum_{t,f}  \left\vert\hat{X}_j(f/K,t)-Y_j(f/K,t) \right\vert .
	\end{aligned}
	\end{equation}
	% V1.2

	We perform validation with every two hours of the training data and stop the training progress
	if no validation improvement exists in 20 consecutive epochs. All the models are trained using 
	Adam optimizer~\cite{kingma2014adam:} with a initial learning 
	rate of 0.001 and a dropout rate of 0.1. The learning rate decays 
	every thirty hours of training data with a decay rate of 0.87. 
	The STFT matrices with a FFT size of 32 ms and a hop size of 8 ms are used 
	as the model input. The actual frame length and shift size (in number of the samples) 
	are automatically calculated with the sample rate of the input audio.
	
	%V1.2
	% \subsection{Model Details}
	
	% For an audio which sample rate is 
	% $sr$ ($\lfloor44100/K\rfloor,K\in{1,2,4,8}$), 
	% frame size (fs) for STFT is $\lfloor sr*0.032\rfloor$ and 
	% hop size is $\lfloor sr*0.008 \rfloor$. And the fft 
	% point equals $2^{\lceil log_{2}(fs) \rceil}$. By doing so, 
	% STFT setting become dynamic and signal with different 
	% sample rate will have same frequency resolution.
%	The length of transform is 2048 for audio sampled at 44.1kHz. Audio with other kinds of sample rate have the same frequency resolution as 44.1kHz. For example, 22.05kHz audio is transformed with 1024 fft points.

	% Table generated by Excel2LaTeX from sheet 'Sheet5'
	\begin{table*}[tb]
		\centering
		\scriptsize
		\setlength{\tabcolsep}{4pt}
				\vspace{-20pt}
		\caption{Model comparison in terms of various metrics on musdb18hq test set}
		
		\begin{tabular}{c|ccc|cccccc|c}
			\toprule
			\hline
			\multicolumn{1}{c|}{} & \textbf{GFLOPs} & \textbf{Params (M)} & \textbf{Train (h)} & \textbf{SAR (A)} & \textbf{SAR (V)} & \textbf{SDR(A)} & \textbf{SDR (V)} & \textbf{SIR (A)} & \textbf{SIR (V)} & \textbf{Average} \\
			\hline
			\hline
			\multicolumn{1}{p{6em}|}{UNET-5} & 182.81 & 13.3  & 61    & 14.20  & 4.32  & 14.62 & 3.16  & 20.89 & 12.61 & 11.63 \\
			$+ \text{CWS}_{K=2}$ & 91.90  & 13.3  & 36    & 14.10  & \textbf{4.97} & \textbf{15.19} & \textbf{4.23} & \textbf{21.98} & 11.99 & \textbf{12.08} \\
			$+ \text{CWS}_{K=4}$ & 46.44 & 13.3  & 26    & \textbf{14.23} & \textbf{5.05} & \textbf{15.56} & \textbf{4.35} & \textbf{22.54} & 12.07 & \textbf{12.30} \\
			$+ \text{CWS}_{K=8}$ & 23.71 & 13.3  & 15  & 14.04 & \textbf{4.73} & \textbf{15.72} & \textbf{4.31} & \textbf{22.00} & 11.58 & \textbf{12.06} \\
			\hline
			\hline
			\multicolumn{1}{p{6em}|}{MMDN} & 27.63 & 0.33  & 59    & 13.22 & 3.73  & 14.50  & 3.12  & 21.18 & 11.73 & 11.25 \\
			\multicolumn{1}{p{6em}|}{MDN} & 37.42 & 0.27  & 32    & 13.94 & 3.35  & 13.90  & 2.59  & 19.40  & 10.56 & 10.62\\
			$+ \text{CWS}_{K=2}$ & 19.03 & 0.27  & 27    & \textbf{13.96} & \textbf{4.11} & \textbf{15.60} & \textbf{3.65} & \textbf{21.30} & \textbf{11.35} & \textbf{11.66} \\
			$+ \text{CWS}_{K=4}$ & 9.67  & 0.27  & 26  & \textbf{14.10} & \textbf{4.00} & \textbf{15.28} & \textbf{3.86} & \textbf{20.91} & \textbf{12.03} & \textbf{11.70} \\
			$+ \text{CWS}_{K=8}$ & 5.07  & 0.27  & 10   & \textbf{13.98} & \textbf{3.85} & \textbf{15.67} & \textbf{4.17} & \textbf{20.68} & \textbf{11.75} & \textbf{11.68}\\
			\hline
			\hline        
			\multicolumn{1}{p{6em}|}{UNET-6} & 220.73 & 53    & 73    & 13.34 & 4.45  & 14.42 & 3.28  & 23.14 & 9.52  & 11.36 \\
			$+ \text{CWS}_{K=2}$ & 110.86 & 53    & 33    & \textbf{14.15} & \textbf{4.73} & \textbf{14.62} & \textbf{3.92} & 22.43 & \textbf{11.50} & \textbf{11.89}  \\
			$+ \text{CWS}_{K=4}$ & 55.92 & 53    & 23    & \textbf{14.39} & \textbf{5.22} & \textbf{16.02} & \textbf{4.79} & 22.63 & \textbf{12.10} & \textbf{12.53} \\
			$+ \text{CWS}_{K=8}$ & 28.46 & 53    & 19    & \textbf{14.01} & \textbf{4.86} & \textbf{15.97} & \textbf{4.95} & 22.63 & \textbf{11.46} & \textbf{12.31} \\
			\hline
			\hline
			\multicolumn{1}{p{6em}|}{BD-UNET-6} & 220.73 & 53    & 149   & 13.87 & 4.79  & 15.20  & 3.94  & 22.73 & 11.33 & 11.98 \\
			$+ \text{CWS}_{K=2}$ & 110.86 & 53    & 92    & \textbf{14.24} & \textbf{4.85} & \textbf{15.44} & \textbf{4.34} & \textbf{22.79} & \textbf{12.76} & \textbf{12.40} \\
			$+ \text{CWS}_{K=4}$ & 55.92 & 53    & 64    & \textbf{14.45} & \textbf{5.24} & \textbf{16.49} & \textbf{5.20} & \textbf{23.12} & \textbf{12.99} & \textbf{12.92}  \\
			$+ \text{CWS}_{K=8}$ & 28.46 & 53    & 57    & \textbf{14.33} & \textbf{4.94} & \textbf{16.06} & \textbf{5.08} & \textbf{22.77} & \textbf{12.70} & \textbf{12.65} \\
			\hline
			\bottomrule
			% \bottomrule
			\vspace{-10pt}
		\end{tabular}%
		\label{tab:main_result}%
	\end{table*}%
	
	\section{Experiments}
	In this section, we will first describe the dataset and evaluation metrics used in this paper. 
	The experimental comparison and analysis of the advantage of CWS will then be discussed.
	
	\subsection{Dataset}
	We mainly conduct experiments on the publicly available \textit{musdb18hq} dataset~\cite{musdb18-hq}. It has a training set with 
	100 songs and a test set of 50 songs. We choose 14 songs from the training set 
	as the validation set, the same as the definition in python package \textit{musdb}\footnote{https://github.com/sigsep/sigsep-mus-db}. 
	To explore the limitation of the data, we also trained 
	our model on an internal training set \textit{aslp}. 
	It has additional 617 songs of pure vocal and 1496 songs of pure instrument, which 
	are collected from the internet. Although some of them may not be absolutely clean,
	experiments show that using additional data improves the separation performance. 
	We follow the steps in ~\cite{uhlich2017improving} for 
	data augmentation. During the training stage, we randomly select, chunk, and 
	mix vocal and instruments and multiply two streams with a scaling factor
	randomly sampled between 0.6 and 1.0. All the songs in \textit{musdb18hq} and \textit{aslp} are 
	stereo and the sample rate is 44.1 kHz.  
	
	\subsection{Evaluation Metrics}
	% The SDR was shown to be valid as a global performance measurement. 
	We use \textit{museval}~\cite{stoter20182018} toolkit to compute SDR, SIR, and SAR~\cite{vincent2005blind} 
	metrics for evaluation. In details, we calculate the metrics 
	for all the segments of the song in the test set with a 
	window size of 1s and hop length of 1s, as commonly used in SiSEC 
	2018~\cite{stoter20182018}. We aggregate both the average SDR, SIR, 
	and SAR by frames as the final score of a song, and pick the median 
	value from each song as the final score of test set.
	All our experiments are performed on a single GTX 1080Ti GPU. For fair comparison, we report some other metrics, e.g., parameter number 
	and training time, as shown in Table~\ref{tab:main_result}. We also use Giga Floating Point 
	Operations (GFLOPs) to weight the computational cost. 
	The floating operation here is measured by a three-second long stereo input.

	% Table generated by Excel2LaTeX from sheet 'Sheet6'
	\begin{table}[t]
		\centering
		\scriptsize
		 \setlength{\belowcaptionskip}{-10pt} 
		 \setlength{\abovecaptionskip}{10pt} 
		 
		\setlength{\tabcolsep}{2pt}
		\caption{Comparison with the state-of-the-art results}
		\begin{tabular}{ccccc}
			\toprule
			Model & \tabincell{c}{Params\\(M)}  & \tabincell{c}{Extra\\Data}  & \tabincell{c}{SDR (A)\\(dB)} & \tabincell{c}{SDR (V)\\(dB)}  \\
			\midrule
			MMDenseNet~\cite{takahashi2017multi}& 0.33  & \checkmark & 15.41 & 3.87 \\
			BLSTM ~\cite{uhlich2017improving} & 30.03 & \checkmark & 14.51 & 3.43 \\
			MMDenseLSTM~\cite{takahashi2018mmdenselstm} & 1.22 & \checkmark & 16.40 & 4.94 \\
			Spleeter-2stem~\cite{hennequin2019spleeter} & 19.6 & \checkmark & 12.88 & 4.72 \\
			MDN & 0.27  & $\times$    & 13.90 & 2.59 \\
			MDN$_{K=8}$ & 0.27  & $\times$    & \textbf{15.67} & \textbf{4.17} \\ 
			UNET-5$_{K=8}$ & 13.3 & $\times$ & 15.72 & 4.31 \\
			BD-UNET-6$_{K=4}$ &53 &\checkmark & \textbf{16.49} & \textbf{5.20} \\ 
			\bottomrule
			\vspace{-20pt}
		\end{tabular}%
		\label{tab:Compare with MMDenseNet with larger dataset}%

	\end{table}%
	\subsection{Result Comparison}
	
	%We perform experiments on unet without CWS input (SUB-NON) and with CWS input (SUB-K). Except for CWS properties, all the experiments have the same setting. We calculate the total loss of validation set every 1.5h of training data. If the validation loss doesn't drop significantly for twenty times or start to raise. The training process will be stopped and the best model will be saved. Fig~\ref{fig.1} shows the training curve and validation loss curve of these four experiments. By comparing the training loss curve, we can see SUB-4 take less time than
	
	The result is shown in Table~\ref{tab:main_result}. 
%	\footnote{The audio samples are avaiable \href{https://ranchofromxgd.github.io/Channel-wise-Subband-Input/}{here}} is shown in Table~\ref{tab:main_result}. 
%	We also offer a demo page \href{https://ranchofromxgd.github.io/Channel-wise-Subband-Input/}{here}. 
	Here we name MMDN and MDN as 
	abbreviation of MMDenseNet and MDenseNet. UNET-$N$ denotes scale $N$ UNet and the 
	prefix BD means the model is trained with extra internal \textit{aslp} dataset. $A$ stands for accompaniment and $V$ stands for vocal.
	Value $K$ stands for total subband number in CWS, as shown in Fig~\ref{fig:ana-syn}.  
	%At first we perform experiments on Unet with a scale of 5. 
	
	In general, the result shows that the CWS input can considerably improve the performance.  
	All the models with CWS surpass the models without CWS on the average SAR, SDR, and SIR by a 
	large margin. Since the computational cost drops drastically with the increase of $K$, 
	the model with a higher $K$ value will converge more quickly. This is beneficial when 
	the dataset is huge.
%	 In that case, a model with CWS can save a great amount of time on experiments or evaluation.  
	 Besides, a higher $K$ will lead to a smaller feature map. 
%	 because the size of the frequency axis will decrease $K$ fold during training. 
	 This can save a lot of memory 
	during training and evaluation, making the model and training process more flexible and
	easier to deploy. 
	
	%V1.2
	As can be seen from Table~\ref{tab:main_result}, splitting 4 bands usually has the 
	best average score on all the evaluation metrics. The average performances of 
	MDN, UNET-5, UNET-6 and BD-UNET-6 increase by 5.7\%, 10.1\%, 10.2\%, and 7.8\%, when using
	the $\text{CWS}_{K=4}$ input. Although $\text{CWS}_{K=4}$ outperforms $\text{CWS}_{K=8}$ by 
	1.5\%, it takes more time, i.e., 38.5\% for model training. 
	Comparing with the model without CWS, $\text{CWS}_{K=8}$ increases the performance 
	by 6.8\% and costs only 31.8\% of the original training time. 
	Moreover, UNET-5/6 and MDN with $\text{CWS}_{K=8}$ achieve 
	the best average SDR, which is valid as a global performance measurement~\cite{vincent2005blind}. 
	Thus in practice, $CWS_{K=8}$ may be the most effective one because it can yield 
	comparable results in a shorter training time. $\text{CWS}_{K=2}$ scenario might be the 
	least preferred setting but still it has contributions to the final score.
	
	It's also worth to mention that $\text{MDN}_{K=8}$ surpasses the 
	performance of MMDenseNet in~\cite{takahashi2018mmdenselstm} even with fewer parameters, 
	far less training data and shorter training time, as shown in Table~\ref{tab:Compare with MMDenseNet with larger dataset}.
%	As was compared in Table ~\ref{tab:Compare with MMDenseNet with larger dataset}. 
	The training set only has 84 songs, but $\text{MDN}_{K=8}$ is still able to exceed the performance of the model trained with a 
	larger dataset. Moreover, our training time might be much shorter. In ~\cite{takahashi2017multi}, a single MDenseNet 
	trained on DSD100~\cite{liutkus2017the} dataset, which is comprised of 100 songs, will take 37 hours for each instrument to train. MMDenseNet trained with extra data 
	will cost more than that time. By contrast, our model only takes 9.7 hours to train. All the evidence strongly demonstrates 
	the advantage of using CWS as model input. Audio samples and codes are available online: \href{https://haoheliu.github.io/Channel-wise-Subband-Input/}{https://haoheliu.github.io/Channel-wise-Subband-Input/}.
	
%	\vspace{-10pt}
	\section{Conclusions}
	
	We present an alternative structure of input feature, namely channel-wise subband (CWS) 
	for VAS model in frequency domain, in order to handle the high computational
	cost and limitation of the conventional CNNs in high-resolution MSS tasks. 
	It overcomes the limitation of the widely used full-band approach 
	%that kernels are shared in all the input field 
	and enables the model to learn weight independently in each subband. 
	Experimental results show that the proposed CWS improve the separation 
	performance and reduce the computational cost significantly. On the public \textit{musdb18hq} 
	dataset, the MDenseNet with 8-bands CWS input exceeds original MDenseNet 
	by 1.67 dB on average SDR of the voice and accompaniment.

%	 As for the configuration of $K$ in CWS, 
%	 generally CWS $K=4$ can get the best overall performance. 
%	 With a little degradation, CWS $K=8$ can achieve comparable 
%	 results in a much shorter time. In some experiments CWS 
%	 $K=8$ even outperform $K=4$ on SDR. Although CWS $K=2$ 
%	 can benefit the training, but it's not good enough 
%	 comparing with $K=4$ or $K=8$, so it's not recommended 
%	 to use this setting. Also we should note that CWS can 
%	 be used on any CNN and frequency-based model. 
	
	%\clearpage
	
	\bibliographystyle{IEEEtran}

	\bibliography{liuhaohe_bib}
	
\end{document}